\documentstyle[12pt,equation]{article}
\advance\voffset by -1.0 cm
\begin{document}
\begin{flushright}
BU-TH-93/1\\
hep-ph/9308239
\end{flushright}
\def\gamgam{\mbox{$\gamma \gamma $}}
\def\ub{\underbar}
\def\qqbar{\mbox{$q \bar{q}$}}
\def\bbbar{\mbox{$b \bar{b}$}}
\def\zzstar{\mbox{$ZZ^*$}}
\def\wwstar{\mbox{$WW^*$}}
\def\wwbar{\mbox{$W^+W^-$}}
\def\ZZ{\mbox{$ZZ$}}
\def\ccbar{\mbox{$c \bar{c}$}}
\def\ttbar{\mbox{$t \bar{t}$}}
\def\llbar{\mbox{$l^+ l^- $}}
\def\ffbar{\mbox{$f \bar f $}}
\def\nunubar{\mbox{$\nu \bar{\nu} $}}
\def\tautaubar{\mbox{$\tau^+ \tau^-$}}
\def\mb{\mbox{$m_b$}}
\def\mh{\mbox{$m_H$}}
\def\mt{\mbox{$m_t$}}
\def\mz{\mbox{$m_Z$}}
\def\mw{\mbox{$m_W$}}
\def\mmu{\mbox{$m_\mu$}}
\def\mpi{\mbox{$m_\pi$}}
\def\mc{\mbox{$m_c$}}
\def\mgut{\mbox{$m_{GUT}$}}
\def\mpl{\mbox{$m_{Pl}$}}
\def\mv{\mbox{$m_V$}}
\def\be{\begin{equation}}
\def\ee{\end{equation}}
\def\bea{\begin{eqnarray}}
\def\eea{\end{eqnarray}}
\def\pbarp{$ \bar p p$}
\def\eplem{\mbox{$e^+ e^- $}}
\def\alphas{\mbox{$ \alpha_s$ }}
\def\rts{\mbox{$ \sqrt{s} $ }}
\def\sigh{ \hat \sigma}
\noindent
\begin{center}
  \begin{Large}
  \begin{bf}
    Higgs Search : Present and Future.\footnote{Talk presented at the
X th DAE symposium, Bombay, December 1992}
\\
  \end{bf}
  \end{Large}
  \vspace{1cm}
  {\it R.M. Godbole\\
    Department of Physics, Univeristy of Bombay, Vidyanagari,\\
    Santacruz(E), Bombay, 400 098, India.}
 \vspace{1cm}
\end{center}
\begin{abstract}
 In this talk I review theoretical bounds on  mass of the
Higgs scalar in  the Standard Model(SM) and then summarise
current experimental limits from the LEP experiments. Following
this I discuss the search strategies for the SM Higgs at  LEP
200 and the TeV energy \eplem\ colliders which are under discussion.
This will be followed by a summary of the Higgs search
potential of the pp supercolliders such as SSC/LHC. I then close
with a brief discussion of a `Dark Higgs' whose dominant decay
modes are into invisible channels.
\end{abstract}

\section*{1)Introduction}
\setcounter{equation}{0}
In spite of the spectacular confirmation of various predictions and
features of the Standard Model (SM), including effects of radiative
corrections, at LEP~\cite{rolandi}  the discovery of the as yet
missing top quark and  Higgs boson is essential for the complete
vindication of the theoretical formulation of the SM. Hence Higgs search
has formed (and will continue to form) an important part of the physics
programme at the current(future) accelerators. Since the presence of Higgs
in the SM is intimately related to the question of giving masses to the
fermions and the gauge bosons, even the failure to find a Higgs boson may
shed light on the mass generation and symmetry breaking mechanism. Various
extensions of the SM almost always enlarge the Higgs sector, but almost
always there is one scalar in these theories which mimics, more or less,
the couplings of a SM Higgs.

 In this talk I will concentrate mainly on the SM Higgs. I will first
briefly review, in section (2), the theoretical `bounds' on the Higgs mass
\mh\ followed  in section (3) by a summary of information about the
branching fractions of  Higgs in  SM into various relevant channels.
Then in the next section I will state current bounds on \mh\ from
LEP data and discuss search strategies/discovery limits at future
\eplem\ colliders: LEP 200 and the TeV energy ($\leq 0.5\ {\rm TeV})$  \eplem\
colliders under planning currently. This will be followed in section (5)
by a discussion of the search possibilities offered by future colliders
like SSC/LHC for the Higgs, with an emphasis on the \gamgam\
signal in the intermediate Higgs mass range: $\mz < \mh < 2\mw, $ the
four lepton signal for the heavier Higgs as well as  use of  forward
jet tagging to isolate  $qq \rightarrow qqH \rightarrow qqWW $
contributions to the WW signal for  Higgs. This technique can prove
useful in the investigations of a strongly interacting vector boson sector
should one exist; a clear non--standard feature.  Following this in
section (6) I will discuss yet another non--SM feature; {\it viz.} the
possibility of a `Dark Higgs' where the dominant decay  mode of the
lightest scalar is into invisible channels. This can affect current
LEP limits on \mh. Then I will end with some conclusions.

\section*{2) SM Higgs: Theoretical mass bounds}
\renewcommand{\theequation}{2.\arabic{equation}}
\setcounter{equation}{0}
As is well known in  SM  couplings of Higgs to matter and gauge
fields are completely predicted~\cite{gunion} but as far as the mass
is concerned there exist only bounds. What is definitely known
is that Higgs can not be too heavy or perturbative theory
breaks down~\cite{sher}. For $\mh\ \geq O(1 {\rm TeV}) $ the
perturbative $VV \rightarrow VV $ scattering amplitude for $V =
Z/W$ violates unitarity~\cite{thacker}. More serious is the fact
that the electroweak theory is not asymptotically free in the
Higgs sector. Thus the self coupling $\lambda$ blows up in
renormalisation group improved perturbation theory. The energy
scale at which $\lambda$ blows up is called the Landau pole. In
addition the  self coupling $\lambda$ increases with $m_H^2$.
Hence the position of the Landau pole itself depends on \mh.
Demanding that  SM with an elementary Higgs scalar should
make sense upto an energy scale $\Lambda$, i.e., the Landau pole
should lie beyond $\Lambda$, then gives an upper bound on SM~\cite{beg}.
\begin{figure}[hbt]
\vspace{8cm}
\caption{Bounds on \mh\ and \mt\ from Landau pole and vacuum stability
\protect\cite{lindner}.}
\label{bounds}
\end{figure}
Fig. \ref{bounds}, taken from~\cite{lindner}, shows the bound obtained
from such an analysis, for different values of $\Lambda$. As we can see for
a light Higgs $\mh\ \leq 180-200$, GeV the perturbative regime is valid
upto $\Lambda \simeq \mgut$ or $\mpl$. As \mh\ increases this region of
validity decreases till finally for $\mh\ \simeq 1$ TeV the theory is
valid only upto $\Lambda \simeq 1$ TeV.

This limit on \mh\ can be understood in a simplified manner~\cite{sher} as
follows. The renormalisation group equation for the quartic coupling
$\lambda$, in the limit of neglecting the gauge and Yukawa couplings, becomes:
\be
{{d\lambda(t)}\over {dt}} = {{3}\over{4 \pi^2}} \lambda^2(t)
\label{rge}
\ee
where $t = ln ({\Lambda}/v)$, where $v$ is the vacuum expectation value and
$\Lambda$ the scale where $\lambda$ is evaluated. Of course this equation
can be solved only when some normalisation condition for $\lambda$ at
scale $v$ is provided. This is chosen to be
\bea
\lambda \equiv \lambda(v) & = & \sqrt{2} G_F m_H^2\\
                        v & = & (2 \sqrt{2} G_F)^{1/2}.
\label{lambda}
\eea
The coefficient of $3/4\pi^2$ in eq. (\ref{rge}) is the lowest order
expression for the $\beta$ -- function which is obtained from the
one--loop corrections to the quartic coupling in the $\lambda
(\phi^\dagger \phi)^2 $ theory. Solving eq. (\ref{rge}) we get
\be
\lambda(t) = {{1}\over{1-3\lambda t /(4\pi^2)}}.
\label{landaupole}
\ee
Then the Landau pole is avoided upto a scale $\Lambda$ provided,
\be
\frac {3} {4 \pi^2} \lambda t =\frac {3} {4 \pi^2} \sqrt{2} {G_F} {m_H^2}
{ln(\Lambda /v)} < 1
\ee
This gives,
\be
\mh \leq \frac {893 {\rm GeV}} {\sqrt{ln (\Lambda /v)}}
\ee
or $\mh\ < 144, 165, 675 $ GeV  for $\Lambda = 10^{19}, 10^{15}
10^{3}$, respectively. The
bounds shown in fig. \ref{bounds} are of course obtained from the full
renormalisation group analysis as the Yukawa couplings, particularly the   top
Yukawa coupling, are non--negligible.

The obvious question is of course validity of a perturbative
analysis of the $\beta $ functions near the Landau pole. However, recent
lattice calculations~\cite{lattice} confirm this bound and conclude that
\be
\mh\ < (8-10) \mw \simeq 600-800 {\rm GeV}.
\ee
Thus, if the SM is to be internally consistent, an upper bound $\sim 1$
TeV  exists on the mass of Higgs--scalar. Hence a search strategy for
SM--Higgs should cover the mass range upto 1 TeV.

There is yet another consideration which leads to bounds on \mh\ {\it
viz.}, considerations of vacuum stability~\cite{sher,chanowitz,linde,weinberg}.
This essentially demands that (i) the one loop corrected scalar potential
$V(\phi)$  has a minimum at $\phi = v$ so as to have spontaneous symmetry
breakdown and further (ii) $V(\phi) \rightarrow \infty$ as $|\phi|
\rightarrow \infty$ so that the Hamiltonian is bounded from below.
At the tree level the scalar potential is,
\be
V(\phi) = - \mu^2 |\phi|^2 +  {{\mu^2}\over {2 v^2}} |\phi|^4 .
\ee
Quantum corrections, at one loop level~\cite{colwei} give us,
\be V(\phi) = - \mu^2 |\phi|^2 + {{\mu^2}\over{2 v^2}} |\phi|^4 +
\gamma |\phi|^4 {\bf \bigg[} ln{{|\phi|^4}\over{v^2}} - {{1}\over{2}} {\bf
\bigg]}
\label{oneloop}
\ee
where
\be
\gamma = {{3 \sum_{vectors} m_v^4 + \sum_{scalars} m_s^4
           - 4 \sum_{fermions} m_f^4}\over {64 \pi^2 v^4}}
\label{gamma}
\ee
The first condition mentioned above would give a lower limit on \mh\  if
the fermions are light so that their contribution to $\gamma$ in eq.
\ref{gamma} can be neglected; {\it e.g. } if $\mt\ < 80 $ GeV, the
requirement (i) above gives $\mh\ > 7 $ GeV~\cite{linde,weinberg}. However
in view of the current bounds on the top mass~\cite{top1} this limit is
by now void. The second requirement really means that $\gamma$ in eq.
\ref{gamma} should not become too negative. This means that \mh\ should
increase with \mt~\cite{chanowitz}. Again at large $|\phi|$ again the one
loop results of eq. \ref{oneloop} can not be valid. The large logarithms
have to be resummed and then a bound on \mh\ has to be obtained. The bound
so obtained in ref.~\cite{lindner} is the one shown in fig. \ref{bounds}.

The above bounds strictly apply only to SM, {\it i.e.} a simple scalar
sector with a single Higgs doublet. In case of a more complicated Higgs
sector these bounds refer to some average mass. This clearly means that
the lower bounds on \mh\ given in fig. \ref{bounds} can be avoided, but not
the upper bounds. Hence if the SM is correct we expect to find
either a Higgs below 1 TeV or some evidence for new physics beyond the
SM or occurence of the onset of new perturbative regime. So in principle,
a comprehensive discussion of search strategies for Higgs at current and
future colliders should include the latter two possibilities as well. Out
of the possible extensions of the SM, supersymmetry~\cite{susy} is perhaps
the most attractive as well as predictive one. These theories have an
extended scalar sector. Search strategies for a supersymmetric Higgs will
be discussed separately at the symposium~\cite{tata}. The possibility of
strongly interacting Higgs sector~\cite{strongww} will not be, however,
covered in much detail here.

\section*{3) Production modes and decays of the Higgs}
\renewcommand{\theequation}{3.\arabic{equation}}
\setcounter{equation}{0}
As said earlier, as a result of Higgs mechanism for spontaneous
symmetry breakdown  Higgs couplings to  fermions and gauge bosons are
completely fixed~\cite{gunion} and are proportional to their masses. Hence,
for a given \mh, the largest branching ratio is into the heaviest
fermion--antifermion pair or the heavy gauge bososn pair. In view of the
above, the CDF limit on \mt~\cite{top1} and  LEP  limits on
\mh~\cite{smlep}, the dominant decay modes for Higgs are
\begin{enumerate}
\item $H \rightarrow \bbbar $ for $\mh < \mz$,

\item $H \rightarrow \bbbar, H \rightarrow \zzstar, H \rightarrow \wwstar $
for Higgs in the intermediate mass range $\mz < \mh < 2\mw $,

\item $H \rightarrow VV$ for $\mh > 2\mv$.
\end{enumerate}
Calculations of the total width and branching ratios of the SM Higgs were
done from scratch in ref.~\cite{lhc} by Kunszt and Stirling.
\begin{figure}[hbt]
\vspace{6cm}
\caption{Decay branching ratios for SM Higgs for  $\mh < 2 \mz$ (a,b) and
$\mh > 2 \mz$ (c) for $\mt = 90$ GeV\protect\cite{lhc1}.}
\label{decays}
\end{figure}
Fig. \ref{decays} shows the decay branching ratios into fermion and gluon
pairs (fig. \ref{decays}(a)) and into a pair of electroweak gauge bosons
(fig. \ref{decays}(b)) for the mass range $\mh < 2 \mz$ as well as for the
dominant channels for $\mh > 2 \mz$ (fig. \ref{decays}(c))  with $\mt = 90
$ GeV.  These branching ratios are not very sensitive to the top mass
apart from the position of  the $\ttbar $ threshold.

In the case of a very light Higgs boson,  which has already been ruled out
by LEP, the following points should be kept in mind:
\begin{enumerate}
\item For $\mh < 2 \mmu$, an ultra light Higgs boson can decay
even outside the detector or c$\tau_H \simeq$ at least a few cm.
\item For $2 \mpi < \mh < 2 \mc $ the calculations of the different
branching ratios have some theoretical uncertainties.
\item For $\mh > 2 \mb $ of course $H \rightarrow \bbbar $ is the dominant
mode, but BR $(H \rightarrow \tau^+ \tau^-) $ is $\simeq 8 \% $. The last
can be seen from fig. \ref{decays}(a).
\end{enumerate}

In the case of the heavier Higgs boson for which the decay branching
ratios have been presented in figs. \ref{decays}(a,b), the dominant
decay mode of Higgs into a \bbbar\ pair suffers from enormous QCD
backgrounds and hence the rare decay mode $H \rightarrow \gamgam
$ plays an important role in the search strategies of such a
Higgs, particularly at a Hadron collider, by providing a much
cleaner final state. The rare decays like $H \rightarrow
\gamgam\ $ and $H \rightarrow gg $ take place via loops and the
loop contribution will be dominated by  gauge boson and heavy
fermions. Hence these rare decay modes are also good
pointers to new physics. The effect of the running of the b
quark mass, has been included in the calculations~\cite{lhc1} and
it reduces the partial width for the $H \rightarrow
\bbbar $ considerably and  hence increases the branching ratios into
channels like $H \rightarrow \gamgam, H \rightarrow gg$ and $H \rightarrow
\tau^+ \tau^- $.

For the intermediate mass Higgs, the channels involving at least one real
vector boson open up and begin to dominate the decay branching width. In fig.
\ref{decays}(b) the dip in the branching ratio in the \zzstar\ channel,
around $2 \mw$, corresponds  to the threshold for the WW channel. Fig.
\ref{decays}(c) shows the dominant branching ratios for $\mh > 2 \mz$. One
can see from there complete domination of  H decays by  gauge
boson and \ttbar\ channels.

\begin{figure}[hbt]
\vspace{4cm}
\caption{Total width $\Gamma^{tot}_H$ as a function of \mh. The
solid(dashed) line corresponds to $\mt = 90(100)$ GeV\protect\cite{lhc1}.}
\label{width}
\end{figure}
Fig. \ref{width} shows the variation of the total Higgs decay width
$\Gamma^{tot}_H$ with \mh\ for two different values of \mt. Again we see
that the dependence of $\Gamma_H^{tot} $ on \mt\ is marginal.
The width rises sharply
once the VV(V=W/Z) channel opens up. For the superheavy Higgs ($\mh >
700-800$ GeV) the total width becomes comparable to the Higgs mass
itself. The calculations of~\cite{lhc1} of the Higgs decay width
and branching ratios have been updated to include the effects of
the  electroweak and QCD radiative corrections~\cite{nlc2}. None
of these change the trends discussed above and amount to a few
\% in most cases.

Just as the dominant decay modes of Higgs are  into the heaviest fermion
pair accessible for a given Higgs mass \mh\ or into a gauge boson
pair, the dominant Higgs production mode at a given collider are also
controlled by  same large couplings. Over most of the Higgs
mass range of interest and the current or planned \eplem\ collider centre of
mass energies, Higgs production is dominantly through the large
HVV coupling and possible processes are shown
\begin{figure}[hbt]
\vspace{6cm}
\caption{Different production modes for a Higgs at an \eplem\ collider.}
\label{feyn1}
\end{figure}
in fig. \ref{feyn1}.
Over most of the mass range of the
Higgs, for $\mt \simeq 100 $ GeV, the production at Hadron colliders takes
place via  gg fusion. Different possible modes of single Higgs
production at a Hadron collider are depicted
\begin{figure}[hbt]
\vspace{3.5cm}
\caption{Different production modes for a Higgs at a hadron collider.}
\label{feyn2}
\end{figure}
in fig. \ref{feyn2}.

{}From above discussions it is clear that the efficacy of the Higgs
search will be decided by the ability to reject against \bbbar\
backgrounds or to use the rare decay modes for a Higgs lighter than $2
\mv$ and by the ability of isolating $VV $ signal due to Higgs
production in the case of heavier Higgses. In the latter case one would
like to be able to distinguish between the $VV$ pairs coming from Higgs
production from gluon fusion(at a Hadron collider) and its subsequent decays
from those due to Higgs production via $VV $ fusion  shown in
fig. \ref{feyn2}.
Furthermore, a discrimination of the $VV$ production due to a strongly
interacting Higgs sector from the above signal, is also desirable.
The details of the expected cross--sections for the signal and the
backgounds at different colliders will be discussed in the next sections.
First I will discuss current limits on the SM Higgs from LEP, followed
by a discussion of search strategies of the Higgs at the
supercolliders~\cite{super}: at a TeV energy \eplem\
collider~\cite{super1,nlc} as well as at the pp
supercolliders~\cite{lhc,super2}.

\renewcommand{\theequation}{4.\arabic{equation}}
\section*{4) Search for the SM Higgs at \eplem\ colliders}
\setcounter{equation}{0}
\subsection*{4a) Higgs search at LEP 100}
\eplem\ colliders are ideally suited for Higgs search due to the
clean environment they offer and the precise knowledge of the initial
energy.  Both these are a considerable advantage over the
hadron colliders, in view of the discussion of section 2.  One
of the production modes for Higgs at an \eplem\ collider is
the Bjorken process~\cite{bj} shown in fig. \ref{feyn1}. This
is the dominant production mode at low \mh\ $(\mh < 70$ GeV), at
the Z pole, which is the energy at which LEP--100 has so far
collected data. Before the LEP collider went into action a
variety of different processes and experiments had provided
limits on \mh \footnote{For a good summary of these mass limits,
see {\it e.g.} ref.~\cite{lepreport}.}. However, almost all these
limits suffered to some extent from theoretical  uncertainties
such as long distance effects in the production of Higgs or in the
calculation of $H \rightarrow \mu^+ \mu^-$ branching ratio when
$\mh < $ few GeV. Hence the limits obtained by LEP, which are
almost free of these uncertainties, are crucially important.

The production process at LEP--100~\cite{bj} is,
\bea
e^+ e^- \rightarrow Z & \rightarrow & Z^* H \nonumber \\
                      & \rightarrow & f \bar{f} H.
\label{brems}
\eea
This will give rise to different final states due to different decays of
the $Z^* $ and $H $. The off shell $Z^* $ decays into a $q \bar{q}$  pair
in about 70\% of the cases, into a $\nunubar$ pair in about 20\% cases and
into a \llbar\ pair remaining $\simeq 10\%$ of the times. In the final
case of course a sum over all the three lepton types ($l = e,\mu,\tau$) is
implied. However, normally only the final states involving $e/\mu $ are
useful.  The dominant final states of Higgs decays will of course
depend on Higgs mass \mh.

The initial suggestion by Bjorken~\cite{bj} was to look for $Z^*$ decaying
into a \llbar\  (with $l = \mu$) final state, regardless of the H decay
and study the distribution in the invariant mass $m_{\mu^- \mu^+}$. The
production of Higgs via the process of eq. \ref{brems} is then
signalled by a peak in this distribution. Fig.
\begin{figure}[hbt]
\vspace{5.5cm}
\caption{BR $(Z \rightarrow H \mu^+ \mu^-)$ (left axis) and total number of
hadronic Z decays which would correspond to 3 $H \ffbar (f = \nu/e/\mu)$
events at the Z pole (right axis) as a function of $m_H$\protect\cite{grivaz}.}
\label{brmumu}
\end{figure}
\ref{brmumu} taken from Ref.~\cite{grivaz}, shows BR $(Z \rightarrow
 H \mu^+ \mu^-)$ (left axis) and total number of
hadronic Z decays which would correspond to 3 $H \ffbar\ (f = \nu/e/\mu)$
events at the Z pole (right axis) as a function of \mh.
{}From this figure, one can see how only with 25,000 hadronic Z decays per
experiment, the nonobservation of a signal in the 1989 LEP data~\cite{LEP89}
could rule out Higgs upto $\mh \simeq 25$ GeV. The very steep dependence
of the BR $(Z \rightarrow H \mu^+ \mu^-) $ on \mh\, however, makes it
clear that further improvements on the limit on \mh, with increasing
number of Z events are slower.

In the search for the light Higgs boson the
four LEP groups (for a good summary see, {\it e.g.}, Ref.~\cite{smlep}),
used not only the final state resulting from $Z^* \rightarrow \mu^+\mu^-$
but also those from $Z^* \rightarrow \eplem$ and $Z^* \rightarrow
\nunubar$ decays. They also used the specific topologies resulting from
different Higgs decays, as allowed in the SM, depending on the value
of \mh  . For an ultra light Higgs boson ($\mh < 2$ GeV) the search is
complicated, in spite of the large production rate, due to the large
variety of possible signals and theoretical uncertainties in the Higgs decay
branching ratios. These somewhat model dependent analyses~\cite{smlep}
excluded a light SM Higgs. OPAL has further excluded a light Higgs upto a
mass of 11 GeV, even in a decay mode independent search~\cite{OPAL}.

For a heavier Higgs boson ($>$ a few GeV),
the process of eq. \ref{brems} gives rise to  following final states
in decreasing order of abundance (for $\mh > 2\ \mb $):
\begin{enumerate}
\item[(i)] ($Z^* \rightarrow hadrons$)($H \rightarrow hadrons$)
\item[(ii)] ($Z^* \rightarrow \nunubar$) ($H \rightarrow hadrons$)
\item[(iii)] ($Z^* \rightarrow \tau^+ \tau^-$) ($H \rightarrow hadrons$) +
       ($Z^* \rightarrow hadrons$) ($H \rightarrow \tau^+ \tau^-$)
\item[(iv)] ($Z^* \rightarrow \eplem/\mu^+ \mu^-$) ($H
\rightarrow hadrons$)
\item[(v)] ($H \rightarrow \tau^+ \tau^-$) $[(Z^* \rightarrow \nunubar), (Z^*
        \rightarrow \llbar)]$
\end{enumerate}
The relative abundances of these various final states, {\it e.g.} for
$\mh = 55 $ GeV, are 64\%, 18\%, 9\%, 6\% and 3\% respectively.
However, due to the large QCD backgrounds, the first channel is quite
ineffective.  The second channel is the most useful as it gives rise to
an acoplanar pair of hadronic jets which can be easily discriminated
against the Z decay background. The high efficacy of this channel at LEP
100 is due to the absence of \ttbar\  production  at this energy. The
final states involving taus are less clear and hence have lower search
efficiency. Therefore the final state in (iv) with somewhat lower rates
but cleaner signature serves better.  This final state, however, suffers
from an irreducible background from $\eplem \rightarrow \qqbar\ \llbar$,
where  the $\qqbar\ (\llbar)$ comes from a photon radiated off a final
state lepton (a final state quark). The bounds on Higgs mass given by
the different LEP groups are given in Table 1~\cite{grivaz}.

\vspace*{6mm}
\noindent
{\bf Table 1:} Lower bounds on the SM Higgs mass from the four LEP
experiments, from ref.~\cite{grivaz}
\begin{center}
\begin{tabular}{|c||c|c|c|c|}
\hline
& ALEPH & DELPHI & L3 & OPAL\\
\hline
$m_H(GeV)$ & 56.3& 47 & 52 & 52.6\\
\hline
\end{tabular}
\end{center}
\vspace*{6mm}

Since the total number of hadronic Z  events collected by all
the four groups by now is $\simeq 2453\ K$~\cite{grivaz}, we
see from fig. \ref{brmumu} that for all the four groups combined
together one would expect $\simeq 3.50 $ H\ffbar\ ($ f =
\nu,e,\mu)$ events. This means that a combined 95 \% confidence
level  limit from all the four groups currently  is $\mh > 60$
GeV. This indicates that no LEP group indvidually will be able
to do better than this limit.  Further progress must come from
the higher energy \eplem\ colliders {\it viz.} LEP 200, the
next linear collider (NLC) and the pp supercolliders.

\subsection*{4b) Higgs search at LEP 200}
  LEP 200 will not run at a fixed energy and is expected to have $175 <
\sqrt{s} < 190$ GeV. The production process is similar to that of eq.
\ref{brems} but the role of the real and the virtual Z's have got
interchanged. The process is
\be
e^+ e^- \rightarrow Z^*  \rightarrow Z +  H
\label{brems1}
\ee
and is shown in fig. \ref{feyn1}. This process has
appreciable cross--sections for $\mh < (\rts - \mz) $.
\begin{figure}[hbt]
\vspace{12cm}
\caption{Higgs production cross-section via the bremsstrahlung process
along with different background processes and their cross--sections for
\protect\rts\ = 175 and 190 GeV \protect\cite{janot}.}
\label{allpro1}
\end{figure}
Figure \ref{allpro1} taken from~\cite{janot} shows the expected cross-section
for Higgs production for different values of \mh\ for
$ 175 < \rts < 190$ GeV.  Again various final
states  possible due to different decays of the Z and H
are the same as discussed in sec. (4a).  The search strategy now
uses the fact that the final state contains a real Z. However, the
search has to proceed differently than at LEP 100 due to the
rise with increasing energy in the cross-sections of various
t--channel processes which should be contrasted with the $1/s$
behaviour of the cross--section of the process of eq.
\ref{brems1}. There exist different t--channel processes such as
$\eplem \rightarrow \ZZ, \eplem \rightarrow \wwbar, \eplem
\rightarrow Z \nunubar, \eplem \rightarrow Z \eplem,  \eplem \rightarrow
 e \nu W $
which lead to final states similar to those one will look here for. Fig.
\ref{allpro1} shows expected  cross-sections for these various background
processes as well.  It can be clearly seen from
the figure that the signal lies well below various background processes.

The most difficult region in \mh\ is the case where  Higgs is
degenerate with  Z {\it i.e.} $\mh \simeq \mz$. In this case the process
\be
e^+ e^- \rightarrow ZZ
\label{ZZ}
\ee
constitutes an extremely serious background as the signal to background
ratio is about 5 and the final states extremely similar.

For  values of \mh\ considered here,  $H \rightarrow \bbbar$ is the
dominant decay mode whereas for  Z the \bbbar\ branching ratio is only
$\sim 14 $\%. So concentrating on the \bbbar\ final states and
using  $\mu$ vertex detectors to tag the `b--quark' one can
handle the background. (It should be perhaps mentioned here that
the `b--tagging ' is useful not only in this degenerate Higgs
case but for the full heavy \mh\ range.) Further the angular
distribution of the  fermions can be used~\cite{ku,brown}
to discriminate ZH signal from ZZ one. The conclusion
of both the references was that at
\rts = 190 GeV an integrated luminosity ($\int {\cal L} {\mbox{dt}} $ =
$5 fb^{-1}$ ) will be needed to get a decent signal to backgound ratio.
A recent analysis of ref.~\cite{janot} takes into account realistic
detector simulation and conludes that using all the channels of the
different Z/H decyas {\it viz.} H\nunubar , H\llbar, \tautaubar, \qqbar\ and
4$q$ jets, one could search right upto $\mh = \mz$, with a similar
integrated luminosity.

\begin{figure}[hbt]
\vspace{4cm}
\caption{Minimum integrated luminosity, $\int {\cal L} dt $,
required for Higgs boson discovery for \protect\rts = 175 [190] GeV
is shown in fig. (a) [(b)]\protect\cite{janot}.}
\label{lumino}
\end{figure}
Fig. \ref{lumino} shows the minimum integrated luminosity required for a
Higgs discovery at a fixed \rts\ as a function of \mh. Discovery here is
being defined as a combined signal at the level of 5 events in all the
different channels and five standard devaitions above the expected
background. Essentially it shows that at the lower energy of $\rts = 175$
GeV the maximum sensitivity to \mh\ ($\simeq 80 $ GeV) is already reached
at an integrated luminosity of $150\  pb^{-1}$. If one should however
increase the energy to 240 GeV, the cross--section for the process of eq.
\ref{brems1}  falls off but even then range of \mh\ accessible
in the process grows. As stated
earlier at a given \rts\ the above process has a reach upto $\rts - \mz$.
At \rts = 240 GeV, this limit will be reached with an integrated luminosity of
300 $pb^{-1}$.

\subsection*{4c) Higgs search at the higher energy \eplem\ colliders:}
At higher energy \eplem\ collider  Higgs prodcution can also take place
via the  WW/ZZ fusion~\cite{fusion} processes
shown in fig. \ref{feyn1}  which are given by,
\bea
e^+e^- & \rightarrow \nunubar W W  \rightarrow \nunubar H \\
e^+e^- & \rightarrow \eplem Z Z  \rightarrow \eplem H.
\eea

The cross--section of the fusion process
begins to grow in importance over the
bremsstrahlung process of the  \ref{brems1}, due to the $ln (s/\mh^2) $
dependence of the former as opposed to the $1/s$ dependence of the latter.
This is clearly seen in the figure.
\begin{figure}[hbt]
\vspace{6cm}
\caption{Higgs production cross--section, in fb, in \eplem\ collisions for
the bremstrahlung and the fusion processes for $\protect\rts = 500$ GeV, as a
function of \mh\protect\cite{nlc1}.}
\label{nlcf1}
\end{figure}
The figure also shows that for the values of \mh\ and \rts\ under
consideration $(\mh > 60 $ GeV, $\rts < 500 $ GeV),
the WW fusion contribution is comparable
to the beamstrahlung one whereas
 the ZZ fusion gives a contribution which is about ten
times smaller. By now the electroweak radiative corrections to
the cross--sections have been calculated~\cite{kniehl}. The QED
corrections are indeed large and come dominantly from the photon
radiation from the initial $e^+/e^-$, whereas the weak
corrections are only at the level of a few \% ($-7\% < \Delta < 4\%$).
On the whole, the radiative corrections are well under control
and well understood.

The real advantage of the high energy linear \eplem\ colliders
is to be able to explore the intermediate mass range for the
Higgs in great detail. As is clear from fig. \ref{decays} (c),
for the relevant \mh\ values the dominant decays are into the
$V^*V^*/VV^*/VV$ where V is W/Z boson. As the discussions in the
last sections show, LEP 200 would certainly probe region upto
$\mh < \mz$. The region $\mh \simeq \mz$ requires special
attention.  For
the case of a Higgs degenerate with the Z, the
characteristic nature of the missing mass spectrum of the signal
$\eplem \rightarrow ZX $, X=H can be used effectively to
separate the signal from the \ZZ\  background~\cite{cheung}. The use
of \bbbar\ final state and flavour tagging using vertex
detectors to identify the b's, also helps reduce the background from \ZZ\
production. For $100 < \mh < 140 $ GeV, the worst background is
single Z production but can again be handled by flavour
tagging. For  $140 < \mh < 160 $ GeV, the dominant background is
from
\be
\eplem \rightarrow \ Z W W^* \rightarrow Z W (W^* \rightarrow q
\bar q'),
\ee
which is again managable as it is suppressed by  electroweak
couplings. Beyond $\mh = 160$ GeV the channel containing three
real vector bosons in the final state open up. Now for a
realisitic estimate of the discovery potential one has to look
at the variety of background processes which can give rise to
similar final states, containing two or more real vector bosons. A
detailed analysis of these backgounds has been
performed~\cite{barger,nlc3}. In these cases the invariant mass
distribution for the VV pair can be used to remove the
background. The conclusion of these studies is that a Higgs
boson upto 0.35 TeV can be observed at an \eplem\ collider with
$\rts = 500 $ GeV, with 10 $fb^{-1}$ luminosity.

One more aim of these \eplem\ colliders will be to use their
cleaner environment (modulo the potential problems that may be
caused by the hadronic backgrounds~\cite{dg} induced by the
beamstrahlung~\cite{beam} photons) to make detailed studies of
the properties of the Higgs, once it is found. Independent of
Higgs decay, the monoenergetic nature of the recoiling Z boson
produced in the reaction of eq. \ref{brems1} can be used to
determine \mh\ once a Higgs signal is seen. In this case the
accuracy of Higgs mass determination is limited by
smearing of the incident electron beam energy caused by the
phenomenon of bemastrahlung~\cite{beam}. A study of the angular
distribution of Higgs produced in the bremsstrahlung process of
eq. \ref{brems1} can yield information about the spin of the
Higgs. Radiative corrections to the angular distributions are
also well under control~\cite{kniehl}.  The strength of the HVV
coupling can be determined from  the production cross--sections. As far the
$Hf\bar f$ couplings are concerned, only their relative strength
w.r.t the HVV couplings can be extracted from the decay
branching ratio measurements. A determination of the $Ht\bar t$
coupling is possible by studying the production of a light Higgs
via  bremsstrahlung off  t($\bar t$)quark~\cite{djozer} or
a light Higgs decay to a \ttbar\ pairs~\cite{hagiwara} as shown
in fig.\ref{feyn1} {\it viz.},
\bea
\eplem &\rightarrow t \bar t H \\
\eplem & \rightarrow t \bar t Z.
\label{higtdk}
\eea
But these processes are useful only in a limited range of
Higgs mass.

{}From the above discussion we can conclude that, \eplem\
colliders with $\rts > 300$ GeV are ideally suited for the
discovery as well as a detailed study of the properties of a SM
Higgs in the intermediate mass range and above.

\section*{5) Higgs search at Hadron colliders:}
\renewcommand{\theequation}{5.\arabic{equation}}
\setcounter{equation}{0}
\subsection*{5a) Higgs search for $\mh > 0.6 - 0.8$ TeV}
Eventhough, the high energy \eplem\ colliders are the ideal
place to look for an intermedaite mass Higgs and
construction of such  linear \eplem\ colliders does seem
possible, they still lie in somewhat distant future~\cite{super1,nlc}.
One of the major goals of the pp supercolliders like the LHC/SSC
that are being planned is the hunt for  Higgs. The expected
cross--sections for different Higgs production processes of
fig.\ref{feyn2}, at the LHC pp collider($\rts = 16 $ TeV) are
shown in
\begin{figure}[hbt]
\vspace{6cm}
\caption{Expected cross--sections for the Higgs production at
the LHC for different production processes of fig.
\protect\ref{feyn2}\protect\cite{lhc1}.}
\label{higxsect}
\end{figure}
fig. \ref{higxsect}. The uncertainties in these cross--sections
are much higher than the corresponding predictions at an \eplem\
collider and they come from uncertainties in the knowledge of
the structure functions as well as the size of the higher order
corrections. For the gg fusion mechanism, which dominates the
production  upto $\mh \simeq 0.8$ TeV, the QCD
corrections~\cite{gghigh} increase the cross--section by as much
as 80\% for both the LHC and SSC energies. This is shown in fig.
\begin{figure}[hbt]
\vspace{4.5cm}
\caption{Higher order QCD corrections to the $gg \rightarrow H$
production, taken from\protect\cite{djouadi}.}
\label{gghighx}
\end{figure}
\ref{gghighx} taken from~\cite{djouadi}.  The figure clearly shows that the QCD
corrections are much higher than the structure function
uncertainties.

This increase is a welcome news as can be seen from the fig.
\begin{figure}[hbt]
\vspace{7cm}
\caption{Cross--sections for different relevant processes at
hadronic colliders from the current $p \bar p$ colliders to
supercolliders \protect\cite{denegri}.}
\label{allpro}
\end{figure}
\ref{allpro} taken from~\cite{denegri}. The Higgs search does
indeed need good search strategies to boost the signal above all
the other, much more abundant, processes. The strategy of course
depends crucially on Higgs mass. The information on
different branching ratios of the Higgs, shown in fig.
\ref{decays}, can be translated into different  effective search
channels for it as shown in fig.
\begin{figure}[hbt]
\vspace{5cm}
\caption{Signatures for Higgs at LHC \protect\cite{denegri}.}
\label{lhcsig}
\end{figure}
\ref{lhcsig}, taken from~\cite{denegri}.
We see from this figure that for the mass range $\mz < \mh < 2
\mz $, the best signal is obtained by using the rare decay mode
$H \rightarrow \gamma \gamma$~\cite{gamgam}. The dominant decay
mode $H \rightarrow \bbbar$ suffers from the enormous QCD
backgrounds shown in fig. \ref{allpro}.

The rates for the process $pp \rightarrow HX \rightarrow \gamma
\gamma X $ are indeed very small, but the signal is observable
over the QCD continuum diphoton background~\cite{seez} in the
$m_{\gamma \gamma}$ distribution.  However, this demands a very
good $m_{\gamma \gamma} $ resolution (to better than 1\%).
A major source of
another reducible background to the signal is  hadron
misidentification and hence very good $\pi/\gamma$ separation
is needed to achieve the rejection factors $\sim 10^8 $ that are
required.The expected number of the signal events and the
intrinsic QCD diphoton background are shown in Table 2.
The numbers in Table 2 are for $m_H = 80-150$ GeV and $\int
{\cal L } {\mbox dt} \simeq 10^{5} pb^{-1}$.
{}From this one can conclude that the $\gamma \gamma $ channel
can make the discovery of Higgs in the intermediate mass
range at a hadronic collider a possibility.

\vspace*{6mm}
\noindent
{\bf Table 2:} The expected number of the $pp \rightarrow HX \rightarrow \gamma
\gamma X $ signal events and the intrinsic bakground at  LHC~\cite{lhc1}.
\begin{center}
\begin{tabular}{|c||c|c|c|c|}
\hline
$m_H$(GeV)& $\Delta$m(GeV) & Signal &Background & $S/\sqrt({B})$\\
\hline
80 & 1.0& 570 &11800 & 5.2\\
\hline
100 & 1.5& 1180& 13700& 10.1 \\
\hline
150& 2.0& 830 & 5600 & 11.1\\
\hline
\end{tabular}
\end{center}
\vspace*{6mm}

However, a word of caution has to be added here. In the above
studies the QCD induced diphoton production has been calculated
using only the Born level box diagrams for the $q \bar q (gg)
\rightarrow \gamma \gamma$. Recent results from CDF~\cite{naba}
show that the box contribution underestimates their measured
$\gamma \gamma $ cross--sections by about a factor 5 and even
the inclusion of the higher order correction still
gives a discrepancy of a factor $\sim 3$ between theory and
experiment. Of course, CDF has had to use very loose cuts in
order to retain a measurable signal and possibility of a
contamination of the signal by misidentified jets can not be
ruled out. In the SSC/LHC environment much stricter cuts
will be allowed which can be used to avoid such contamination.
This disagreement also underscores the importance of a good
knowledge of the structure functions for these predictions.
In view of the crucial role played by this channel in
the Higgs search, this issue needs to be studied further more
carefully.

Another possibility~\cite{lhc2} for the intermediate masss Higgs is to look
for associate production of the W and H followed by the decay of
Higgs in two photons :
\be
pp \rightarrow W H X \rightarrow (W\rightarrow l \nu) (H
\rightarrow \gamma \gamma ) X
\ee
Again the rates are very small but the background from $pp
\rightarrow W \gamma \gamma $ is even smaller. The numbers given
in Table 3 show that such a study seems feasible. Again the
numbers are for an integrated luminosity of $10^5 pb^{-1}$ for LHC.
The WHX signal can also be augmented by $t\bar t H X $ channel
where the H decays into the \gamgam\ channel~\cite{tthx}. The two
channels together do provide a possibility for the intermediate
mass Higgs detection~\cite{denegri}.

\noindent{\bf Table 3:} The expected number of events for the
signal ($pp \rightarrow HW \rightarrow (H \rightarrow \gamgam)(W
\rightarrow l \nu) $ and background expected at  LHC~\cite{lhc1})

\vspace{6mm}
\begin{center}
\begin{tabular}{|c||c|c|c|c|}
\hline
$m_H$(GeV)& Signal &\multicolumn{3}{c|}{Background} \\
\hline
&&irreducible&reducible&total\\
\hline
75 & 17& 6 & 1 & 7\\
\hline
100 & 22& 3& 1& 4 \\
\hline
130& 18& 2 & $<1$ & 3\\
\hline
\end{tabular}
\end{center}
\vspace*{6mm}

\begin{figure}[hbt]
\vspace{6cm}
\caption{Higgs signal corresponding to $\mh = 600$ and 800 GeV,
expected at the LHC for an integrated luminosity of $10^{5}$
$pb^{-1}$, taken from ref. \protect\cite{denegri}. The solid line shows
the expected background.}
\label{golden}
\end{figure}

For $\mh > \mz$, the best signal is via the process~\cite{fourl},
\be
pp \rightarrow HX \rightarrow ZZX \rightarrow \llbar \llbar X
\ee
Fig. \ref{golden} shows the expected signal in this channel for
$\mh = 600$ and 800 GeV, along with the expected background.
Thus we see that it is possible to device cuts to get a
significant signal, above the background, in the four lepton
channel upto $\mh = 800 $ GeV. From fig. \ref{width} it can be
seen that $\mh \sim \Gamma_H$, and the concept of  Higgs as a
fundamental particle does not make much sense. This value is
also at the upper end of the limits on  Higgs mass implied by
the theoretical arguments mentioned in Section 2. Thus we see
that the pp colliders will be able to look for a standard model
Higgs over the whole practical Higgs mass range, beyond the one
that is accessible to LEP 200. The pp collider SSC has a
discovery potential similar to that of the LHC only the
luminosity required would be a factor 10 smaller due to the
higher energy of the SSC~\cite{sschiggs}.
\subsection*{5b) Heavy Higgs ($m_H > 0.6-0.8 $ TeV)}

For a heavy Higgs, even for large values of the top mass VV
production mechanism of for Higgs given by,
\be
qq \rightarrow qqVV \rightarrow qqH \rightarrow qqVV
\label{vv}
\ee
dominates over the gluon fusion mechanism
\be
gg \rightarrow H \rightarrow VV.
\label{gg}
\ee
If $\mt = 100(150)$ GeV the former begins to dominate if $\mh >
0.6(0.8)$ TeV. For such large values of \mh, the possibility of
strongly interacting Higgs sector can not be ruled out. This
means that VV scattering could produce the same final states
as due to Higgs production and decay. Eventhough, the Higgs
sector does not become strongly interacting, it would be still
necessary to separate the two production processes of the Higgs
given in eqs. \ref{vv} and \ref{gg}. The suggestion to use the high
energy, `forward' jets to `tag' the VV fusion events was put
forward~\cite{kahnww} quite some back. Recently this was
analysed for the WW decay mode of H at the LHC
workshop~\cite{lhc3} and then in more detail for both qqWW and
qqZZ final states~\cite{zeppenfeld}. The basic idea here is that
the forward jets in the process of eq. \ref{vv} have high
energy, high rapidity and $p_T \sim O(m_W)$. The backgrounds are due
to the QCD processes (i) $q \bar q(gg) \rightarrow t \bar t, q \bar q(gg)
\rightarrow t\bar t g$, (ii)$q \bar q \rightarrow W^+ W^- g$ and
of course (iii) the WW scattering.  All these are about three to
four orders of magintude higher than the signal. However
judicious choice of cuts on the rapidities and $p_T$  of the
jets gives about 25\% efficiency for the $qq \rightarrow qqWW $
channel  and roughly 70\% efficiency for the $qq \rightarrow
qqZZ $. The feasibility of such tagging is very important from
the point of vies of analysing the strongly interacting $W/Z^s$,
should they exist.
\section*{6) Dark Higgs.}
\renewcommand{\theequation}{6.\arabic{equation}}
\setcounter{equation}{0}

So far we discussed only the SM Higgs search where the dominant
decay modes were always into the heaviest fermion--antifermion
pair or gauge boson pair. In some extensions of the SM there
exists a light scalar but its dominant decay modes are
`invisible'. Here we are talking about a heavy Higgs ($>$ a few
GeV) which is `invisible'. Some examples are
\begin{enumerate}
\item Certain parameter space in Supersymmetric theories where
the dominant decay modes of the lightest Higgs are into
neutralinos~\cite{griest}.
\item Majoran models~\cite{majoran} with spontaneously broken lepton number
which have Higgs decaying dominantly into a Majoran pair
\item Models where the lightest scalar decays into a pair of
Goldstone bosons~\cite{dark}.
\end{enumerate}
Griest and Haber had studied the implications of this for
Higgs search at LEP, in the first case. Recently Djouadi\cite{djouad1} and
collaborators analysed this issue in view of the LEP constraints
on the SUSY parameter space. Their conclusion is that the
parameter space which corresponds to a light, `invisible' Higgs,
will have other light sparticles {\it i.e.}, neutralinos and
charginos, which should have been seen at LEP-100 or should be
seen at LEP 200.
The issue of the `invisible' Higgs in the Majoran models was
taken up recently~\cite{anjan1,anjan2}. This work demonstrated
that it is possible to construct models with a global symmetry
such that Higgs will have invisible decays without causing the Z
to have large invisible width. A similar analysis in the context
of models with spontaneously broken R--parity was also done~\cite{Jose}.
Now the signal due to the bremsstrahlung process \ref{brems1} at LEP-100
would be
\be
\eplem \rightarrow Z^* \rightarrow \ffbar + {\rm missing energy}.
\ee
The (theoretical) analyses of~\cite{anjan1,Jose} show that
indeed the LEP mass limits on \mh\ can be affected by this and
the issue should be examined carefully.

If such an invisible Higgs is in the intermediate mass range
then it will lead to very characteristic missing energy
signatures at the hadron colliders~\cite{invishy} via the process
\be
pp \rightarrow ZHX; pp \rightarrow WHX.
\label{whx}
\ee
A preliminary analysis of the viability of this signal against
the background from the processes
\be
pp \rightarrow ZZX; pp \rightarrow WZX
\ee
followed by the decay of Z into \nunubar\ has been done in first
of~\cite{invishy}.  It seems indeed possible to separate the signal due to
processes of eq. \ref{whx} from the background. Hence the phenomenology of
such an invisible Higgs for different mass ranges and different colliders
is an interesting topic for future investigations.

\section*{7) Conclusions}
Our discussion  can be summarised as follows:
\begin{itemize}
\item If we demand that the Standard Model be consistent upto an
energy scale $\Lambda \sim 1 $ TeV a Higgs scalar must exist
with $ \mh < 600-800 $ GeV or we should find some evidence for new physics
beyond the SM or new perturbative regime at an energy scale
$\sim 1$ TeV.
\item LEP 100 has ruled out a SM Higgs with $\mh < 60 $ GeV.
Further improvements on this limit can now come only from LEP 200.
\item Future \eplem\ colliders with $\rts = 500 $ GeV should be
able to look for the SM Higgs upto $\mh \sim 350 $ GeV and can
in principle afford a detailed determination of its properties
to confirm that it is indeed the SM Higgs, should one be found.
\item Since the \eplem\ colliders still lie somewhat in the
distant future the real Higgs discovery machines are perhaps the $pp$
supercoliders LHC/SSC. These should be able to look for the SM Higgs over
the entire range of interest (upto $\mh = 800 $ GeV) with an integrated
luminosity of $10^5 pb^{-1}$ ($10^4 pb^{-1}$) at the LHC(SSC).
\item The $H \rightarrow \gamgam$ decay mode plays a crucial
role in making the detection of the SM Higgs over the entire
mass range possible.  QCD computation of the diphoton production
seems to underestimate the current CDF data and this issue needs
to be studied carefully before drawing conclusions about the
observability of the SM Higgs in the \gamgam\ channel.
\item There exist models where the dominant decay mode of the
lightest scalar is into invisible channels and which can affect
the current LEP bounds. Phenomenology of such models needs to be
investigated.
\end{itemize}


\newpage
\begin{thebibliography}{99}
\bibitem{rolandi}
L. Rolandi, CERN-PPE/92-175, Talk given at the XXVI th
International Conference on High Energy Physics, Dallas, August (1992).

\bibitem{gunion}
See, for example, J.F. Gunion, H.E. Haber, G.L. Kane and S.
Dawson, `The Higgs Hunters Guide' (Addison-Wesley, 1990)

\bibitem{sher}
See, for example, M. Sher, Phys. Rep.
\underbar{179} (1989) 273, and L. Maini, Rome preprint no. 775
(1991), lectures given at the Cargese Summer School, 1990.

\bibitem{thacker}
B.W. Lee, C. Quigg and H. Thacker, Phys. Rev.
\underbar{D16} (1977) 1519.

\bibitem{beg}
M.A.B. Beg {\it et al.}, Phys. Rev. \underbar{52} (1984) 883;
D.J. Callaway, Nucl. Phys. \underbar{B233} (1984) 189;
R. Dashen and H. Neuberger, Phys. Rev. Lett. \underbar{50}
(1983) 189;
K.S. Babu and E. Ma, Phys. Rev. \underbar{D31} (1984) 2861;
E. Ma, Phys. rev. \underbar{D31} (1985) 322.

\bibitem{lindner}
M. Lindner, Z. Phys. \underbar{C31}(1986)295.

\bibitem{lattice}
A. Hasenfratz, K. Jansen, C. Lang, T. Neuhaus and H. Yoneyama,
Phys. Lett. \underbar{199B}(1987)531;
J. Kuti, L. Liu and Y. Shen, Phys. Rev. Lett. \underbar{61}
1988) 678;
M. L\"uscher and P. Weisz, Nucl. Phys. \underbar{B318} (1989) 705.
\bibitem{chanowitz}
M. Chanowitz, M. Furman and I. Hinchliffe, Phys. Lett.
\underbar{B78} (1978) 285;
N. Cabbibo {et al.}, Nucl. Phys. \underbar{B158} (1979) 295;
R.A. Flores and M. Sher, Phys. Rev. \underbar{D27} (1983) 1679.

\bibitem{linde}
A. D. Linde, Sov. Phys. JETP Lett., \underbar{23} (1976) 64.

\bibitem{weinberg}
S. Weinberg, Phys. Rev. Lett. \underbar{36} (1976) 294.

\bibitem{colwei}
S. Coleman and E. Weinberg, Phys. Rev. \underbar{D7} (1973) 1838.

\bibitem{top1}
See, for example, J. Huth, `Top quark search from CDF',talk
presented at the same conference as in \protect\cite{rolandi}.


\bibitem{susy}
See, for example,
J. Wess and J. Bagger, `Supersymmetry and Supergravity'
(Princeton University Press, New York, NY. 1983) ;
P. West, `Introduction to Supersymmetry and Supergravity',
(World Scientific, Singapore, 1986).

\bibitem{tata}
X. Tata, these proceedings.

\bibitem{strongww}
For a review of developements in this subject, see, for example,
G. Altarelli, CERN-TH 6317(91), talk given at the International
Workshop on Electroweak Physics beyond the Standard Model,
Valencia, Spain (October 1991).

\bibitem{smlep}
For a review, see, fro example, E. Gross and P. Yepes, CERN-PPE/92-153, To
appear in the International Journal of Modern Physics A.

\bibitem{lhc}
G. Jarlskog and D. Rein(eds), Proc. Large Hadron Collider
Workshop, Aachen 1990, Report CERN 90-10, Vols. I and II.

\bibitem{lhc1}
D. Froidevaux, Z. Kunszt and W.J. Stirling {\it et al.}, in Vol. II of ref.
\protect\cite{lhc}.

\bibitem{nlc2}
A. Djouadi, D. Haidt, B. Kniehl. B. Mele and P.M. Zerwas,
Proceedings of the workshop `\eplem\ collisions at 500 GeV, the
physics potential' DESY-Report 92-123, ed. P. Zerwas and
references therein.

\bibitem{super}
See, e.g., Proceedings of the Workshop on Physics at Future
Accelerators, La Thuile, 1987, ed. J.H. Mulvey, CERN Yellow book, 87-07.

\bibitem{super1}
Proceedings of the 1st Workshop on the Japenese Linear
Collider (JLC), KEK, Tsukuba, 1989, ed. S. Kawabata, KEK report 90-2.

\bibitem{nlc}
Proceedings of the workshop in \protect\cite{nlc2}.

\bibitem{super2}
 See, e.g., Proceedings of the Workshop on Experiments,
Detectors and Experimental Areas for the Supercollider, Berkeley, 1987,
eds. R. Donaldson and M. Gilchriese, World Scientific, 1988.

\bibitem{bj}
J.D. Bjorken, Proceedings of Summer Institute on Particle
Physics, SLAC report \underbar {198} 1976;
J. Ellis, M.K. Gaillard and D.V. Nanopoulos, Nucl. Phys.
\underbar{B106} (1976) 292;
B. Ioffe and V. Khoze, Sov. J. Part. Nucl. Phys. \underbar{B9}
1978 50.

\bibitem{lepreport}
P. Franzini {\it et al.} in `` Z Physics at LEP I '', CERN
89--08, Vol. 2, p.59, {\it eds.} G. Altarelli, R. Kleiss and C. Verzegnassi.

\bibitem{grivaz}
J.F. Grivaz, Higgs Boson Searches at LEP I, LAL 92-59.

\bibitem{LEP89}
M. Davier, Proceedings of the $XXVth$ High Energy Physics
Conference, Singapore, Aug. 1989, World Scientific, Ed. K. Phua.

\bibitem{OPAL}
P.D. Acton {\it et al.} (OPAL Coll.), Phys. Lett.
\underbar{268B} (1991) 122.

\bibitem{janot}
P. Janot, `Will a Higgs boson be found at future \eplem\
colliders?', LAL 92-27.
\bibitem{ku}
R. Kleiss, Z. Kunszt and W.J. Stirling, Phys. Lett.
\underbar{242B} (1990) 507.
\bibitem{brown}
N. Brown, Z. Phys. \underbar{C49} (1991) 657.

\bibitem{fusion}
D.R.T. Jones and S.T. Petcov, Phys. Lett. \underbar{84B} (1979) 440.

\bibitem{nlc1}
A. Djouadi, D. Haidt and P.M. Zerwas,
in same proceedings as in \protect\cite{nlc2}

\bibitem{kniehl}
J. Fleischer and F. Jegerlehner, Nucl. Phys. \underbar{B216}
(1983) 469;
A. Denner, B.A. Kniehl and J. K\"ublbeck, in the same
proceedings as in \protect\cite{nlc2} and references therein.

\bibitem{cheung}
K. Cheung, private communication as quoted in \protect\cite{nlc2}

\bibitem{barger} V. Barger, K. Cheung, B.A. Kniehl and R.J.N.
Philips, Phys.  Rev. \underbar{D46} (1992) 3725.

\bibitem{nlc3} P. Grosse Wiesmann, D. Haidt and H.J. Schreiber,
in same proceedings as in \protect\cite{nlc2}.

\bibitem{dg}
M. Drees and R.M. Godbole, in same proceedings as in \protect\cite{nlc2}.

\bibitem{beam}
V.N.~Baier and V.M.~Katkov, Phys. Lett. \underbar{A 25},
492(1967).

\bibitem{djozer}
A. Djouadi, J. Kalinowski and P.M. Zerwas, Z. Phys.
\underbar{C54} (1992) 255.

\bibitem{hagiwara}
K. Hagiwara, H. Murayama and I. Watanabe, Nucl. Phys.
\underbar{B367} (1991) 257.

\bibitem{gghigh} A. Djouadi, M. Spira and P.M. Zerwas, Phys.
Lett. \underbar{264B} (1991) 440;
S. Dawson, Nucl. Phys. \underbar{B359} (1991) 283.

\bibitem{djouadi} A. Djouadi, Higgs bosons at future colliders,
DESY-92-128(1992).

\bibitem{denegri} A. Denegri, Rapporteur talk in \protect\cite{lhc}.

\bibitem{gamgam}
J.F. Gunion, G.L. Kane and J. wudka, Nucl. Phys. \underbar{B299}
(1988) 231.

\bibitem{seez}
C. Seez {\it et al.}, in \protect\cite{lhc}.

\bibitem{naba}
See, for example,
N. Mondal, these proceedings.

\bibitem{lhc2}
R. Kleiss, Z. Kunszt and W.J. Stirling, Phys. Lett.
\underbar{253B} (1991) 269;
M. DiLella {\it et al.} in \protect\cite{lhc}.

\bibitem{tthx}
J.F. Gunion, Phys. Lett. \underbar{253B} (1991) 269;
W.J. Marciano and F.E. Paige, Phys.  Rev. Lett. \underbar{66} (1991) 2433;
Z. Kunszt, Z. Trocsanyi and W.J. Stirling, Phys. Lett.
\underbar{271B}(1991) 247;
A. Ballestrero and E. Maina, Phys. Lett. \underbar{299B} (1993) 312.

\bibitem{fourl}
M.S. Chanowitz and M.K. Gaillard, Nucl Phys. \underbar{B261}
(1985) 379;
R. N. Cahn and M.S. Chanowitz, Phys. Rev. Lett. \underbar{56}
(1986) 1327.

\bibitem{sschiggs}
A large number of detailed studies exist. For a comprehensive
review, see, for example, R. Barnett {\it et al.} and J. F.
Gunion {\it et al.}, Proceedings of the Snowmass Workshop 1988
and 1990.
\bibitem{kahnww}
R. Kleiss and W.J. Stirling, Phys. Lett. \underbar{200B} (1988) 193.

\bibitem{lhc3}
M. Seymore {\it et al.} in \protect\cite{lhc}.

\bibitem{zeppenfeld}
V. Barger, K. Cheung, T. Han, J. Ohnemus and D. Zeppenfeld, Phys. Rev.
\underbar{D44} (1991) 1426;
V. Barger, K. Cheung, T. Han and D. Zeppenfeld, {\it ibid.} 2701.

\bibitem{griest}
K. Griest and H. Haber, Phys. Rev. \underbar{D37} (1988) 719.

\bibitem{majoran}
For an early proposal see,
R.E. Shrock and M. Suzuki, Phys. Lett. \underbar{100B} (1982) 250.

\bibitem{dark}
J.D. Bjorken, `What lies ahead?', SLAC preprint SLAC-PUB-5673.

\bibitem{djouad1}
A. Djouadi, private communication.

\bibitem{anjan1}
A.S. Joshipura and S.D. Rindani, Phys. Rev. Lett. \underbar{69} (1992) 3269.

\bibitem{anjan2}
A.S. Joshipura and J.W.F. Valle, CERN-TH-6652/92 (To appear in
Nucl. Phys. B).

\bibitem{Jose}
J.C. Romao, F. de Campos and J.W.F. Valle,
Phys. Lett. \underbar{292 B} (1992) 329.

\bibitem{invishy}
S.G. Frederikson, N.P. Johnson, G.L. Kane and J.H. Reid,
SSCL-preprint-577(1992);
J.C. Romao, J.L. Diaz-Cruz, F. de Campos and J.W.F. Valle,
FTUV/92-39.

\end{thebibliography}
\end{document}